\newcommand{\beq}{\begin{equation}}
\newcommand{\eeq}{\end{equation}}
\documentclass{ws-procs9x6}

\begin{document}

\title{``Branes'' in lattice $SU(2)$ gluodynamics}

\author{V.I. Zakharov}

\address{Max-Planck Institut f\"ur Physik, \\
F\"ohringer Ring, 6 \\ 
80805 M\"unchen, Germany\\ 
E-mail: xxz@mppmu.mpg.de}


\maketitle

\abstracts{We review lattice evidence for existence of 
thin vortices in the vacuum state of 
$SU(2)$ gluodynamics.  On the average, the 
non-Abelian action density per unit area
is ultraviolet divergent  as $a^{-2}$
where $a$ is the lattice spacing. At small scales, the surface looks very
crumpled so that the corresponding entropy factor appears also ultraviolet
divergent.
The total area, however,  scales in physical units.
The surface is populated by  monopoles which represent a tachyonic mode.
The smallest value of $a$ tested is about $(3~ GeV)^{-1}$.}

\section{Introduction}

Initially, interest to magnetic monopoles in non-Abelian theories 
was motivated by the dual-superconductor model of the  confinement which
assumes monopole condensation, for review see, e.g., 
\cite{review}. More recently, there emerged a somewhat related mechanism of 
condensation of P-vortices, for a review see, e.g., \cite {greensite}.

In this mini-review we also discuss monopoles and vortices. However,
we are not directly interested in the confinement mechanism.
Instead, the emphasis is on the anatomy 
of monopoles and vortices as field-theoretical objects,
that is on their action and entropy. The lattice is a unique means to 
measure action and entropy directly without relying on a particular
model. And in case of the monopoles and vortices the results
turn to be unexpected.

To appreciate the observations,  let us first
remind the reader that there are two typical scales for
the vacuum fluctuations, that is the lattice spacing $a$ and 
$\Lambda_{QCD}$. The lattice spacing provides an ultraviolet cut off
while $\Lambda_{QCD}$ represents an infrared-sensitive scale. 
By $\Lambda_{QCD}$ one can understand
a renorm-invariant quantity:
$$
\Lambda^2_{QCD}~\approx~a^{-2}\big(4\pi b_0\alpha_s(a^2)\big)^{-b_1/b_0}
e^{-1/4\pi b_0\alpha_s(a^2)}~,$$
where $b_0$ and $b_1$ are the first two coefficients of the
$\beta$-function. Hereafter we shall have in mind in fact $SU(2)$
gluodynamics, $b_0=11/24\pi^2,b_1=17/128\pi ^4$. 
Alternatively, by $ \Lambda_{QCD}$ one can understand  
square root from the string tension, $\sigma_{SU(2)}$.

Both $a$ and $\Lambda_{QCD}$ are relevant to vacuum fluctuations. 
Im particular, the vacuum energy density, $\epsilon_{vac}$ is
dominated by the zero-point fluctuations and is 
ultraviolet divergent:
$$
\epsilon_{vac}~\approx~\sum_{{\bf k}}{\omega({\bf k})\over 2}~\sim~a^{-4}~,
$$  
where the summation includes also summation over the degrees of freedom.
On the lattice, $\epsilon_{vac}$ is directly related  to the 
average plaquette action
$\langle P\rangle$:
\beq
\langle 1- P\rangle_{pert}~=~{c_G\over a^4}~~,
\end{equation}
where the coefficient $c_G$ is calculable perturbatively,
for details and references see \cite{gc}. Explicit dependence
on $a$ is typical for perturbative contributions.

As for non-perturbative vacuum fluctuations, one usually thinks in
terms of `bulky' fields of the  size of order $\Lambda^{-1}_{QCD}$.
We would call this picture instanton motivated. In  particular,
for a n instanton of a typical size
\beq
\langle 1- P\rangle_{instanton}~=~{\tilde{c}_G}\Lambda^4_{QCD}~,~
\end{equation}
and one expects similar estimates to be true for a generic non-perturbative
fluctuation.

However, recent lattice data (see \cite{kovalenko} and references therein)
 strongly indicate
that monopoles and vortices represent a new kind of 
fluctuations which are ``fine tuned''. Namely, the density of
{\it non-Abelian} action associated with these fluctuations is
ultraviolet divergent on the presently available lattices.
Thus, the action is sensitive to the ultraviolet scale $a$.
However, the total length of the monopole trajectories, respectively,
total area of the vortices are in physical units. Thus, both scales,
$a$ and $\Lambda_{QCD}$ coexists for the monopoles and vortices.   
This is the phenomenon of fine tuning. Actually, the
fine tuning of the Standard Model is generically of the same type,
as we will explain later. 

Another remarkable feature is that monopoles are associated with
a two-dimensional surface rather than with the whole
of the space. The evidence comes from measurements of
the total density of the monopole clusters \cite{boyko2,chernodub}.
Moreover, strong correlation between the monopoles and 
center vortices was noticed first in Ref \cite{giedt}
for a particular value of $\beta$ and later confirmed
for other lattice spacings \cite{kovalenko}.
The surfaces are
populated by monopoles or, alternatively, one can say that the tachyonic 
mode (monopoles) lives on a two-dimensional sub-manifold of the original
four-dimensional (Euclidean) volume \cite{kovalenko}.
 
In this sense, there observed a kind of branes
in the vacuum state of $SU(2)$ gluodynamics (in the Euclidean space).
It is well known (for a review see, e.g., \cite{savit}) that fluctuations
which appear as topological excitations in one formulation of a theory
can become fundamental entities in a dual formulation of the same theory.
Appearance of D-branes in dual formulations of non-Abelian gauge theories
has been widely discussed recently \cite{maldacena}.  
We are in haste to add, however, that there is no known relation whatsoever 
between the ``branes'' seen on the lattice and branes of string theories.
The search process for the branes on the lattice 
is of pure heuristic nature and,
as a result, interpretation of the lattice data remains mostly an open
question.

The talk is based on the original papers 
\cite{kovalenko,bornyakov,boyko1,boyko2,vz,chernodub}.

\section{Definitions of the monopole trajectories and vortices}
\subsection{Topological defects}

The trajectories and surfaces are defined on the lattice 
as topological defects 
in projected field configurations. Topological defects in gauge theories 
are well known of course and here
we will  mention only a few points. The most famous example seems to be
instantons. The corresponding topological charge is defined as 
\begin{equation}
Q_{top} = \frac{g^2}{32\pi^2}\int G_{\mu\nu}^a G_{\rho\sigma}^a
\epsilon^{\mu\nu\rho\sigma}d^4 r~~,
\end{equation}
where $G^a_{\mu\nu}$ is the non-Abelian field strength tensor, $a$ is
the color index, $a=1,2,3$.
For a field configuration with a non vanishing charge there exists
a non-trivial bound on the action:
\begin{equation}\label{bound}
S_{cl}~ \ge~ |Q_{top}| \cdot \frac{8\pi^2}{g^2}~~.
\end{equation}
Instantons saturate the bound.


If we would restrict ourselves to a $U(1)$ subgroup of the $SU(2)$,
instantons would not appear but instead we could discuss magnetic monopoles.
The topological charge now is given in terms of the magnetic flux:
\begin{equation}\label{magneticcharge}
Q_M~=~{1\over 8\pi}\int {\bf H}\cdot d{\bf s}~~,  
 \end{equation} 
where ${\bf H}$ is the magnetic field and $\int d{\bf s}$ is the integral
over surface of a sphere. Note that 
the magnetic field in (\ref{magneticcharge})
does not include the field of the Dirac string due
to the lattice regularization (for details see, e.g., \cite{coleman,alive}).
For a non-vanishing $Q_M$
the  corresponding magnetic mass diverges in the ultraviolet:
\beq\label{mmass}
M_{mon} = \frac{1}{8\pi}\int\limits^\infty_a \mbox{\boldmath${\rm
H}$}^2 d^3 r \sim \frac{1}{e^2} \int\limits^\infty_a
\frac{d^3r}{r^4} \sim \frac{\rm const}{e^2 a}~, 
\end{equation}
where $a$ is an ultraviolet cut off, the overall constant depends in fact
on the details of the cut off and we kept explicit the factor $1/e^2$ which
is due to the Dirac quantization condition, $Q_M\sim 1/e$.

It is convenient to translate the bound on the mass (\ref{mmass})
into a bound on the action $S_{mon}$ since it is the 
action which controls the probability to find a fluctuation.
The translation is straightforward once we realize that monopoles are 
represented by closed lines, or trajectories of a length $L$. Indeed,
the ultraviolet divergence in the mass, see (\ref{mmass}), implies
that the monopole can be visualized as point like 
while conservation of the magnetic charge means that the 
trajectories are closed. Thus, the monopole action in case of  $U(1)$
gauge theory is bounded as:
\beq\label{maction}
S_{mon}~\ge~{const\over e^2}{L\over a}~~,
\end{equation}
where by $a$ we will understand hereafter the lattice spacing.

It is worth emphasizing that the bound (\ref{maction}) is not valid
if we embed the $U(1)$ into $SU(2)$ and there are indications that any
definition of a ``monopole'' can be realized on a non-Abelian field
configuration with a vanishing action, for details see \cite{alive}. 

Finally, we can consider the $Z_2$ subgroup of the original $SU(2)$.
For the $Z_2$ gauge theory the natural topological excitations
are closed surfaces (for review and further references see, e.g.,
\cite{greensite}). Indeed, in this theory the links can be 
$\pm I$ where  $I$ is the unit matrix. Respectively the plaquettes 
take on values $\pm 1$. Unification of all the negative plaquettes 
is a closed surface and the action is
\beq\label{vortexaction}
S_{vort}~=~const{A\over a^2}~~,
\end{equation}
where $A$ is the area of the surface.
Again, the infinitely thin vortices are natural
excitations only in case of $Z_2$ gauge theory. 

\subsection{Projected field configurations}

The physical idea behind definition of the lattice monopoles is the
so called Abelian dominance. According to the hypothesis of
the Abelian dominance, it is the Abelian degrees of freedom
which are responsible for the confinement in $SU(2)$ as well.
Since monopoles are natural 
objects only in the  $U(1)$ case one should replace, or project
an original configuration of $SU(2)$ fields into the closest $U(1)$
configuration. If the idea of the Abelian dominance is correct
the effect of the projection is not dramatic.
At the next step, one defines  monopoles
in terms of the projected configuration
as if it were a $U(1)$ theory from the very beginning.
We still have to explain what is understood by the ``closest''
$U(1)$ field configuration.

\begin{figure}[ht]
\centerline{
\psfig{file=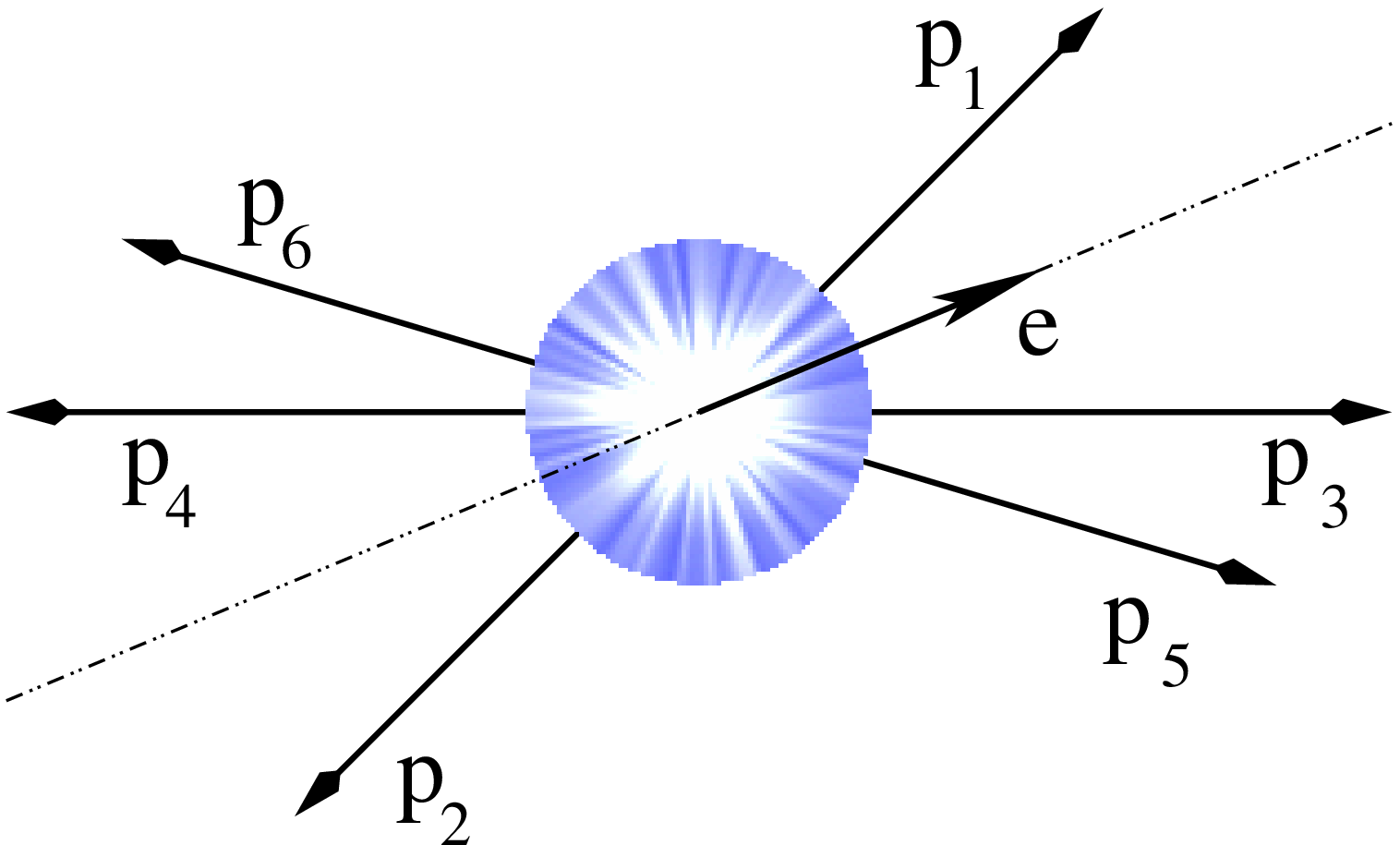,width=0.5\textwidth,silent=}
\hspace{0.05\textwidth}
\psfig{file=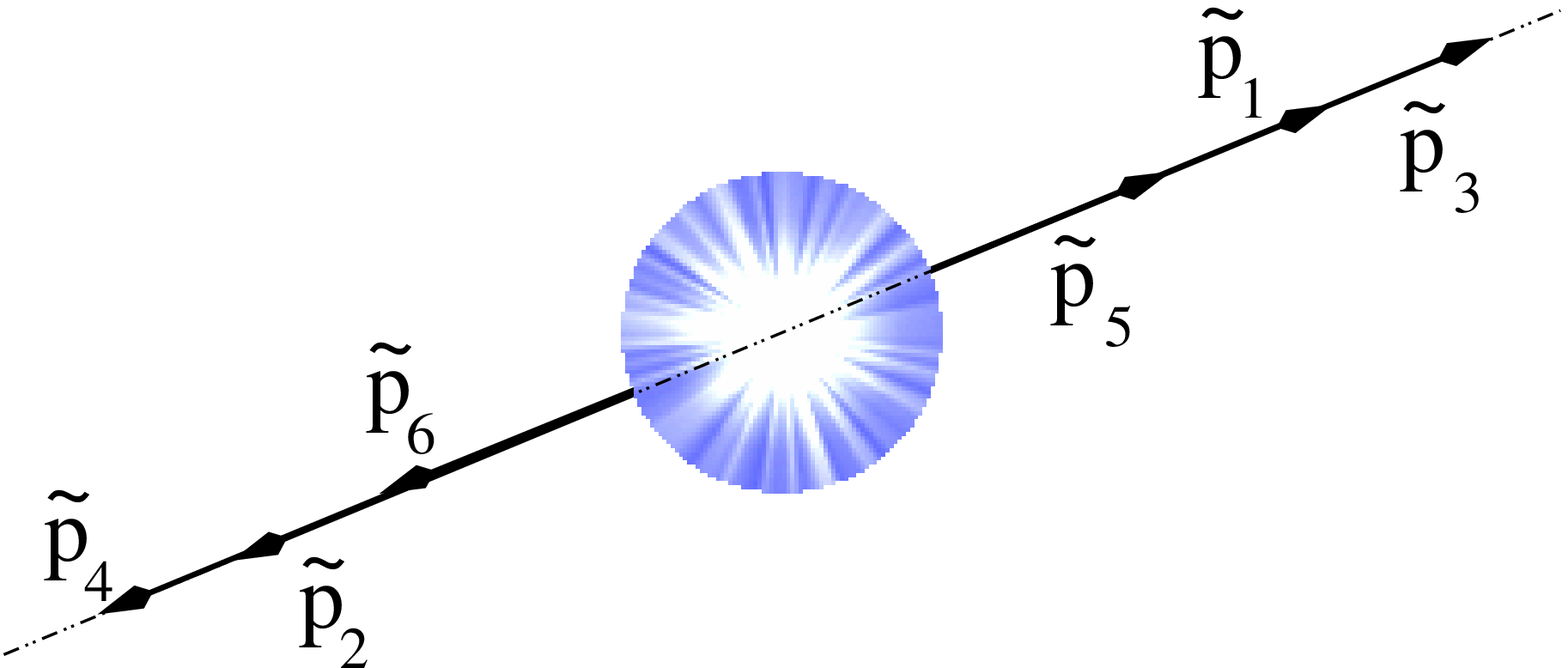,width=0.5\textwidth,silent=}
}

~

\centerline{~~~~a)~~\hspace{0.5\textwidth} ~~~~~b)} \caption{a)~
Particle momenta and choice of the axis ({\bf e}) by maximizing
the sum of momenta projections on the axis; \newline b) the corresponding
collinear momenta closest to the original ones.}
\end{figure}

Let us mention first a simple analogy. Imagine that two jets of particles
are produced in a central collision (see Fig. 1a).
We suspect that it is the properties of the longitudinal components of
the momenta which are most important.
Let us define then the ``closest'' collinear configuration  of
momenta in a the following two steps.
First, using the rotational invariance, choose an axis,
or unit vector {\bf e}, 
in such a way that the sum of moduli of projections of the momenta
on the axis is maximal:
$$ {\rm max} \sum_i
\mid\mbox{\boldmath${\rm p}$}_i \mbox{\boldmath${\rm e}$}\mid~~.
$$
Second, replace the original momenta by their projections to the 
axis defined in this way:
$$
\mbox{\boldmath${\rm p}$}_i \to \mbox{\boldmath${\rm \tilde{p}}$}_i \equiv
\mbox{\boldmath${\rm e}$}
\mid\mbox{\boldmath${\rm p}$}_i \mbox{\boldmath${\rm e}$}\mid {\rm
sign} \; (\mbox{\boldmath${\rm p}$}_i \mbox{\boldmath${\rm e}$})
\;\; .
$$
It is natural to call the momenta constructed in this way as the
collinear configuration closest to the original one.
(see Fig.~1b).

In case of a gauge theory, our basic object is
the potential $A_\mu^a$, where $a$ is the color index, $a=1,2,3$.
 Let us label $A_\mu^3$
as electromagnetic field and $A_\mu^{1,2}$ as fields of charged 
particles (gluons). As the first step -- in analogy with the
example given above-- let us fix the gauge in a  way which maximizes
the neutral field:
 $$
{\rm max} \sum_{i,\mu} \mid A_\mu^3(x_i)\mid^2~, $$ 
where $x_i$ --
are positions of the lattice sites. As the next step, 
put $A_\mu^{1,2} \equiv 0$.

As a result, we replace the original field configuration of
$A_\mu^{1,2,3}$ by the `closest' Abelian field configuration  $\bar
A_\mu^3$.
Our magnetic  monopoles are then nothing else but the Dirac monopoles
in terms of the projected fields $\bar A_\mu^3$. The Dirac monopoles 
correspond  to singular fields and the corresponding monopole current
can be determined in terms of violations of the Bianchi identities:
\begin{equation}
\partial_\mu \epsilon_{\mu\nu\rho\sigma} \partial_\rho \bar
A_\sigma \equiv j_\nu ~.\label{16}
\end{equation}
Better to say, one uses a lattice analog of Eq. (\ref{16})
so that all the singularities are uniquely determined \cite{degrand}.

The results of the lattice measurements reduce then to a set
of monopole trajectories. First, one generates $SU(2)$ fields.
At this step nothing is used but the original
non-Abelian action of the $ SU(2)$ theory. Then each 
configuration is replaced by its Maximal Abelian projection.
Finally, one determines on the projected fields the
monopole trajectories.The procedure can be iterated for 
various values of the lattice spacing $a$. Data on the monopole trajectories
is the starting point of our analysis.

\section{Notion of fine tuning}
\subsection{Action-entropy balance}

Let us concentrate first on $U(1)$ gauge theory.
The natural topological excitations are then monopoles.
However, the action (\ref{maction}) is ultraviolet divergent
and, at first sight, the monopole contribution is enormously suppressed in
the limit $a\to 0$. This is actually not true. The point is that the entropy 
is also divergent in the limit $a\to 0$ (for a detailed explanations see, e.g.,
\cite{polyakov}). 

Indeed, the entropy factor is given now by the number 
$N_L$ of monopole trajectories
of the same length $L$. This number can be evaluated only upon introduction
of discretization of the space-time. For a hyper-cubic lattice the
number is \cite{shiba}:
\beq \label{entropy}
N_L~=~exp(\ln7\cdot L/a)~~.
\end{equation}
Thus, the probability to find a monopole trajectory of length $L$ is 
proportional to:
\beq\label{probability}
W_L~\sim~exp\big(-{c\over e^2}+``\ln 7''\big)\cdot{L\over a}
\end{equation}
where we put $\ln 7$ in quotation marks since Eq (\ref{entropy})
does not account for neighbors (numerically, though, the 
effect of neighbors is small \cite{shiba}).

The probability (\ref{probability}) is a function
of the electric charge alone. In particular, if $e^2$ is equal to
its critical value,
\beq
e^2_{crit}~\approx~{\ln 7\over c }~~,
\end{equation}
then any length $L$ is allowed and the monopoles condense.

Eq. (\ref{probability}) demonstrates also that in the limit $a\to 0$, 
generally speaking,
monopoles are either very rare or too common, depending on the
sign of the difference in the 
exponential. Only a very narrow band of values of $e^2$,
\beq\label{phys}
e^2_{crit}~-~\tilde{m}\cdot a~\le~
e^2_{phys}~\le~e^2_{crit}~+~\tilde{m}\cdot a~~,
\end{equation}
can be called physical. Here $\tilde{m}$ is a constant of dimension
of mass. Indeed, only in this case the monopole condensate is controlled
by the scale of $\tilde{m}$ independent of $a$.

\subsection{Supercritical phase}

Keeping in mind application to non-Abelian theories, the most interesting
case is 
$$ e^2-e^2_{crit}~<~0,~~\big(e^2_{crit}-e^2\big)~\ll~1~.$$
In the language of the percolation theory
(for exposition and references see, e.g., \cite{grimmelt}), 
this choice corresponds 
to the supercritical phase.
In this phase there always exists a single percolating cluster which percolates
through the whole volume (of the lattice). Moreover, the probability 
$\theta(link)$
for a particular link to belong to the percolating cluster is small:
$$\theta(link)~\sim~\big(e^2_{crit}-e^2\big)^{\alpha}~~,$$
where the critical exponent $\alpha$ is positive, $0<\alpha<1$.

\subsection{Relation to the fine tuning of the Higgs mass}

It is worth emphasizing that the fine tuning (\ref{phys}) is actually  
similar to the fine tuning of the Standard Model. Indeed, 
for the Higgs particle one has:
\beq\label{higgs}
m^2_{Higgs}~=~M^2_{rad}~-~M^2_0~~,
\end{equation}
where $~M^2_{rad}$ is the radiative correction and 
$~M^2_0$ is a counter term. Both $~M^2_{rad}$ and $~M^2_0$ are
quadratically divergent in the ultraviolet while $m^2_{Higgs}$ 
is presumably independent of the ultraviolet cut off.

One can readily figure out that there should be a connection between
(\ref{phys}) and (\ref{higgs}). Indeed, in the field theoretical language,
$m^2_{Higgs}=0$  manifests `beginning' of the condensation of the
Higgs field. Similarly, the condition $e^2=e^2_{crit}$ 
ensures condensation of the monopoles. 

On a more technical level, similarity of the conditions $m^2_{Higgs}=0$ and
$e^2=e^2_{crit}$ is established in the following way
(see, e.g., \cite{polyakov,ambjorn}).
One starts with the action of a free particle in the polymer representation:
\beq\label{polymer}
S_{polym}~=~M(a)\cdot L~~,
\end{equation}
Moreover, one can define the propagator a la Feynman:
\beq\label{propagator}
\tilde{D}(x, x^\prime) = \sum_{paths}\exp(-S_{polym}(x,
x^\prime))~~.
\end{equation}
and demonstrate that (\ref{propagator})
indeed reproduces--up to an overall constant--propagator of a free
scalar particle. However, the propagating mass, $m_{prop}$ does not
coincide with the `polymer' mass $M(a)$ but is rather given by:
\beq\label{propagating}
m_{prop}^2 = {8\over a}\big(M(a) -{``\ln 7''\over a}\big)~~.
\end{equation}
Comparing (\ref{propagating}) and ({\ref{higgs})
we see that the two expressions coincide with each other
provided that the counter term in Eq. (\ref{higgs}) 
is identified with the $\ln 7$ term in
Eq. (\ref{propagating}). And, indeed, the condition $e^2=e^2_{crit}$
corresponds to $m^2_{Higgs}=0$.

\subsection{Fine tuning of $Z_2$ vortices}

In case of the $Z_2$ gauge theory the topological excitations are
closed vortices which unify all the negative plaquettes.
The vortex tension is ultraviolet divergent and the phase transition
to percolating vortices corresponds again to a fine tuning of
the action and entropy factors.

\section{Fine tuning seen on the lattice}  

\subsection{Fine tuning of the lattice monopoles}

Lattice measurements reveal remarkable scaling properties of the monopole
clusters defined in the maximal Abelian projection (MAP). In particular,
there is always a single percolating monopole cluster, as is expected
in the supercritical phase (see above). Moreover, one can measure the density 
of the monopoles $\rho_{perc}$ in the percolating cluster defined as:
\beq\label{perc}
L_{perc}~\equiv~\rho_{perc}\cdot V_4~~,
\end{equation}
where $V_4$ is the lattice volume, $L_{perc}$ is the lengths of
the percolating cluster in the lattice volume.

The density of the percolating clusters scales (see \cite{mueller}
and references therein):
\beq\label{percolating}
\rho_{perc} \simeq 0.62 \sigma_{SU(2)}^{3/2} \approx
c_{perc}\Lambda_{QCD}^3 ~~,
\end{equation}
where $\sigma_{SU(2)}$ is the string tension in the $SU(2)$ gluodynamics.
Note that Eq. (\ref{percolating}) implies 
that the probability that a given link (on the dual lattice)
belongs to the percolating cluster is proportional to $a^3$ :
\beq\label{theta} 
\theta (link)~\sim~(a\cdot \Lambda_{QCD})^3~.
\end{equation}
What is most remarkable about the probability (\ref{theta}) is
that it perfectly $SU(2)$ invariant. Despite of the fact 
that the definition of the monopoles assumes choosing a particular
$U(1)$ subgroup for the projection and this subgroup is defined 
non-locally, in terms of the whole of the lattice.

From now on, our strategy will be to assume that there are gauge
invariant entities behind the monopoles detected through the MAP
\footnote{In particular, this assumption was formulated
in \cite{teper}. For a discussion of $SU(2)$ invariant monopoles
see \cite{faber,berry}.}.  
The assumption might look too bold and is difficult to justify
on general grounds. But we believe that this could be a right way to make
progress: to accept that observations like (\ref{percolating})
imply that through the maximal Abelian projection we detect gauge
invariant objects. And instead of trying to justify this 
from first principles go ahead
and look for further consequences.

There is another puzzling feature of (\ref{percolating}). Namely,
if we tend $a\to 0$
the same length of the percolating cluster is added up 
from smaller and smaller pieces. As if the local object had physical meaning.
In other words, assuming that the scaling is not accidental
one could have concluded that the monopoles
are point-like (at least, at the presently available lattices)
and are associated therefore with an ultraviolet divergent action!
Unfortunately, the prediction had not been made before the measurements
were done. But, anyhow, the measurements do reveal an action of
order $L/a$, see \cite{bornyakov}
and references therein. The results are reproduced in Fig. 1.

\begin{figure}[ht]
\centerline{\epsfxsize=3.9in\epsfbox{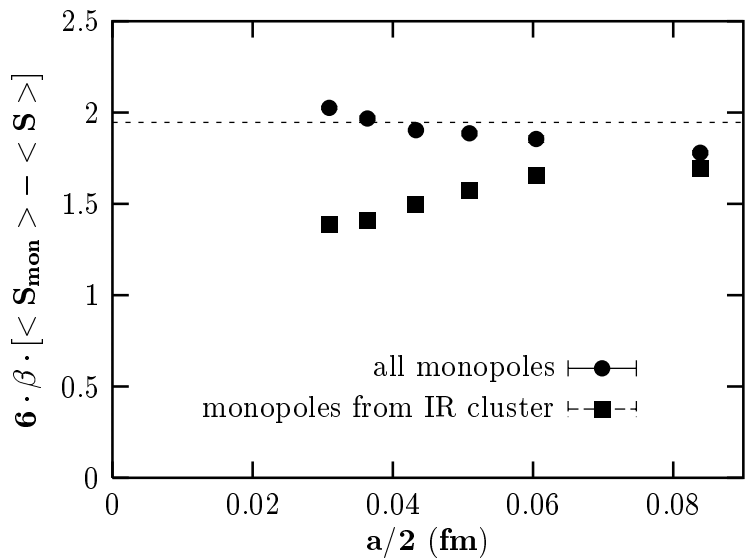}}   
\caption{Excess of the non-Abelian action associated with monopoles
\protect\cite{bornyakov}. 
Squares are for the percolating monopoles and circles are
an average over all the monopoles. The dashed line corresponds to the
monopole mass $\ln 7/a$.\label{nabel}}
\end{figure}

The procedure was to define first the monopoles through the projection
of each configuration of the non-Abelian fields. The monopoles occupy
then centers of certain cubes on the lattice. Then, one
measures the full non-Abelian 
action on the plaquettes closest to the centers of the
cubes occupied by the monopoles, averages over the monopoles and subtracts 
the average over the whole lattice. 

What is to be remembered from the Fig. 1 is that the monopole action 
can be approximated by a constant independent on $a$ and this is
so in the {\it lattice units of action}. The units themselves are
proportional to $a^{-4}$ and singular in the continuum limit. The
reason for using such units is that typical fluctuations on the lattice
are zero-point fluctuations and the corresponding action density is
indeed ultraviolet divergent.


To summarize, the fine tuning of the monopole 
trajectories has been discovered on the lattice: the monopole 
trajectories are associated with singular action and thin
while their length does not depend on the lattice spacing. Moreover, 
the probability
for a link to belong to the percolating cluster vanishes in the limit
$a\to 0$, see (\ref{theta}). This means that in this limit we are exactly 
at the point of the phase transition to the monopole percolation.

\subsection{Fine tuning of vortices on the lattice}

To define  vortices, or two-dimensional surfaces one projects further
the $U(1)$ fields $\bar{A}_{\mu}^3$ into the closest $Z_2$ fields, i.e.
onto the matrices $\pm I$. The surfaces are then unification of
all the negative plaquettes in terms of the projected $Z_2$ fields.
By definition these surfaces are infinitely thin and closed. Their
relevance to confinement has been intensely investigated, see reviews
\cite{greensite} and references therein.

We are interested in the entropy and non-Abelian action 
associated with the surfaces. The results of the measurements 
\cite{kovalenko} are
reproduced in Figs. 2,3. At first sight, there is nothing
dramatic: in both cases we have only weak dependence on $a$.
The `drama' is in the units: the total area per volume
is approximately constant
in physical units while the action density is a constant in lattice
units.  
\begin{figure}[ht]
\centerline{\epsfxsize=3.9in\epsfbox{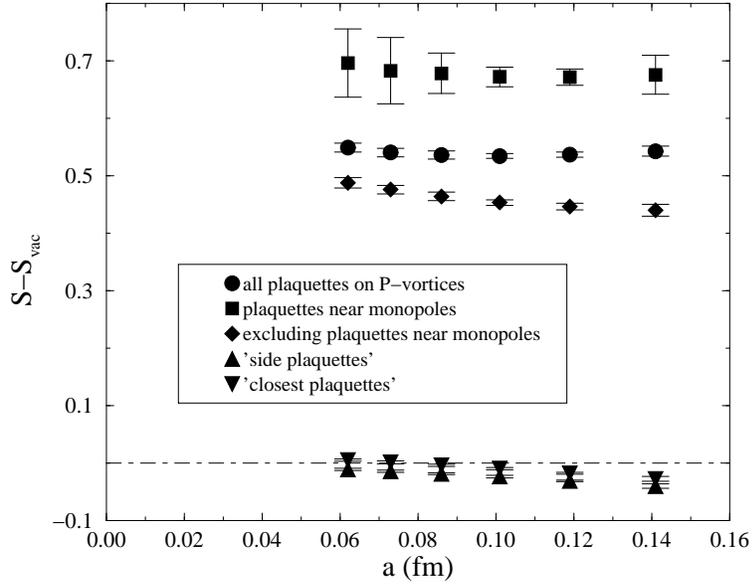}}   
\caption{Excess of the non-Abelian action associated with 
the vortices \protect\cite{kovalenko}. The excess of  non-Abelian action 
is measured separately
on the average over the vortex and on the plaquettes which
simultaneously belong to monopoles. On the neighboring plaquettes
(geometrically, there two different types of them) there
is no excess of the action. \label{actioneps}}
\end{figure}

Thus, the excess of the action associated with the surface is
approximately
\beq
S_{vortex}~\approx~0.5 {A\over a^2}~,
\end{equation}
where $A$ is area and $a$ is the lattice spacing.
While for the total area of the percolating surfaces one finds:
\begin{equation}
A_{vortex} \simeq 4 ({\rm fm})^{-2} V_4~~,
\end{equation}
where $V_4$ is the volume of the lattice.

\begin{figure}[ht]
\centerline{\epsfxsize=3.9in\epsfbox{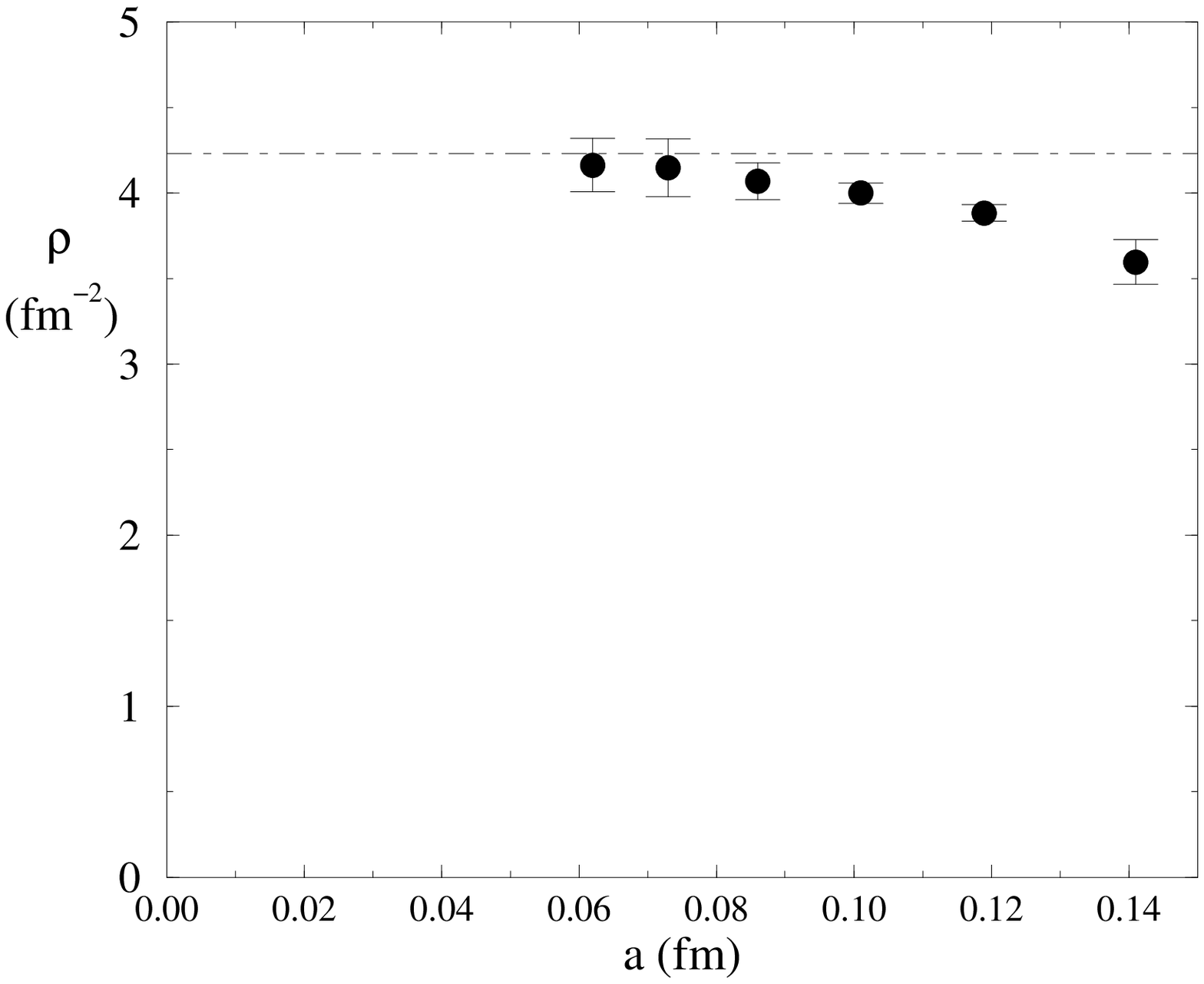}}   
\caption{Scaling of the total area of the vortex \protect\cite{kovalenko}.
\label{rhoeps}}
\end{figure}

Thus, one can say that coexistence of the infrared and ultraviolet scales
in case of the surfaces is directly seen on the lattice.

\subsection{Monopole clusters at short distances}

Taken at face value, the data on monopoles bring us to an
amusing conclusion that monopoles make sense at short distance
and might be treated as point-like particles
whose action is fine tuned to the entropy.
Then one can try to understand the properties of finite monopole clusters
\cite{chernodub,boyko2}.

The simplest vacuum graph for free monopoles is just a closed loop
without self-intersections. This graph corresponds to the the following
spectrum of the clusters in their length $L$:
\beq\label{spectrum}
N(L)~\sim~L^{-3}~~,
\end{equation}
as can be understood by inspecting, e.g., equations in Ref. \cite{polyakov}.
The spectrum (\ref{spectrum})
remains true with account of the Coulomb-like interaction 
as well \cite{chernodub}.
Moreover, the radius of the cluster, $R$ should satisfy the relation:
\beq\label{radius}
R~\sim~\sqrt{L\cdot a}~~.
\end{equation}
Both predictions (\ref{spectrum}) and (\ref{radius}) 
are in perfect agreement with the data \cite{teper,boyko1,boyko2}.

Thus, we can say that the simplest vacuum loop
corresponding  to the monopole field  has been directly observed on 
the lattice.

\subsection{Association of the monopoles with the vortices}

We have just argued that at short distances the monopoles behave as
point-like particles \cite{chernodub,boyko2}. However, viewed from a large
distance, or in the infrared the properties of the cluster change dramatically
as is seen from the data on the total monopole density \cite{boyko2}.
Namely, the full monopole density 
is fitted as \cite{mueller,boyko2}:
 \beq\label{fulldensity}
\rho_{perc}~+~\rho_{fin}~\approx~c_1\Lambda_{QCD}^3~+~
c_2\Lambda_{QCD}^2a^{-1}~,
\end{equation}
where $\rho_{fin}$ is the density of the monopoles
in the finite clusters (defined similar to (\ref{perc})) and
$c_{1,2}$ do not depend on $a$.

Geometrically Eq (\ref{fulldensity}) 
in its generality implies that the monopole trajectory are
associated with a two-dimensional sub-manifold of the whole d=4
space. 
Moreover, the fact that the monopoles spread only over 
vortices whose total area scales like $\Lambda_{QCD}^{-2}$
was first found  empirically \cite{giedt} for one value of $a$ and confirmed
later for the whole range of $a$ available now \cite{kovalenko}.
These vortices are just the vortices discussed in detail above.

\section{Conclusions}

Lattice data strongly indicate that fine tuning is quite common 
phenomenon, at least at presently available lattices.
Namely, both the radiative mass of the monopoles and tension
of P-vortices -- defined in terms of the excess of non-Abelian action--
are ultraviolet divergent. On the other hand, the length of the percolating
monopole cluster and total area of the vortices scale in physical
units.

To estimate how fine the tuning is, one can compare the radiative
mass of the monopole $M(a)_{mon}$ and it free path $L_{free}$ \cite{boyko1}:
\beq\label{tuning}
M_{mon}(a)~>~5 GeV~~,~~L_{free}~\approx~1.6 ~fm~~.
\end{equation}
The free path is defined as the distance measured along the trajectory
between intersections within the percolating cluster. It scales 
in physical units \cite{boyko1}. The radiative mass, $M(a)$ is defined
in terms of the non-Abelian action (see Fig. 2) and $5~GeV$ corresponds to 
the lowest value of the lattice spacing $a$ available now, $a_{min}\approx
(3GeV)^{-1}$. Naively, one could expect $L_{free}\sim (M_{mon}(a))^{-1}$.
In reality it is about 40 times larger.

The most straightforward interpretation of observations like (\ref{tuning})
is a huge cancellation between the action and entropy factors \cite{vz}.
Indeed, one can check that the entropy factor is ultraviolet divergent
and the divergence is similar to the divergence in the action. However,
the cancellation itself is not checked independently to any 
reasonable accuracy. Still, in case of monopoles the `tuned value'
of the mass, $M(a)\sim \ln 7/a$ falls rather close to the data,
see Fig. 2. Also, properties of the percolating monopoles cluster
are similar to the properties of any percolating system in
supercritical phase and very near the phase transition.

We have not discussed interpretation of the data much. 
For a well defined reason: monopoles and vortices are defined
in terms of projected fields and this obscures their relation
to the original theory. Although the properties of the excitations
detected through Abelian and central projections turn to be 
perfectly $SU(2)$ invariant the nature of the monopoles and vortices
remains an open question.

\section*{Acknowledgements}

The author is grateful to the organizers of the workshop,
and especially to Prof. K. Yamawaki for the invitation and hospitality.
The work was partially supported by the grant INTAS-00-00111 and
and by DFG program ``From lattices to phenomenology of hadrons''.

\end{document}